\documentclass{aa}
\usepackage{graphicx}
\usepackage[varg]{txfonts}
\usepackage{cases}
\usepackage{natbib}

\begin{document}

   \title{Spectroscopic observations of a flare-related coronal jet}

   \author{Q. M. Zhang\inst{1,3}, Z. H. Huang\inst{2}, Y. J. Hou\inst{3,4}, D. Li\inst{1,3}, Z. J. Ning\inst{1}, and Z. Wu\inst{2}}

   \institute{Key Laboratory of Dark Matter and Space Astronomy, Purple Mountain Observatory, CAS, Nanjing 210023, PR China \\
              \email{zhangqm@pmo.ac.cn}
              \and
              Shandong Provincial Key Laboratory of Optical Astronomy and Solar-Terrestrial Environment, 
              Institute of Space Sciences, Shandong University, Weihai 264209, PR China \\
              \and
              CAS Key Laboratory of Solar Activity, National Astronomical Observatories, Chinese Academy of Sciences, Beijing 100101, PR China \\
              \and
              School of Astronomy and Space Science, University of Chinese Academy of Sciences, Beijing 100049, PR China \\
              }

   \date{Received; accepted}
    \titlerunning{A flare-related coronal jet}
    \authorrunning{Zhang et al.}

 \abstract
   {Coronal jets are ubiquitous in active regions (ARs) and coronal holes.}
   {In this paper, we study a coronal jet related to a C3.4 circular-ribbon flare in active region 12434 on 2015 October 16.}
   {The flare and jet were observed in ultraviolet (UV) and extreme ultraviolet (EUV) wavelengths by the Atmospheric Imaging Assembly (AIA) on board the Solar Dynamics Observatory (SDO).
   The line-of-sight (LOS) magnetograms of the photosphere were observed by the Helioseismic and Magnetic Imager (HMI) on board SDO.
   The whole event was covered by the Interface Region Imaging Spectrograph (IRIS) during its imaging and spectroscopic observations.
   Soft X-ray (SXR) fluxes of the flare were recorded by the GOES spacecraft. Hard X-ray (HXR) fluxes at 4$-$50 keV were obtained from observations of RHESSI and Fermi.
   Radio dynamic spectra of the flare were recorded by the ground-based stations belonging to the e-Callisto network.}
   {Two minifilaments were located under a 3D fan-spine structure before flare. The flare was generated by the eruption of one filament. The kinetic evolution of the jet was divided into two phases:
   a slow rise phase at a speed of $\sim$131 km s$^{-1}$ and a fast rise phase at a speed of $\sim$363 km s$^{-1}$ in the plane-of-sky.
   The slow rise phase may correspond to the impulsive reconnection at the breakout current sheet. The fast rise phase may correspond to magnetic reconnection at the flare current sheet.
   The transition between the two phases occurred at $\sim$09:00:40 UT. The blueshifted Doppler velocities of the jet in the Si {\sc iv} 1402.80 {\AA} line range from -34 to -120 km s$^{-1}$.
   The accelerated high-energy electrons are composed of three groups. Those propagating upward along open field generate type \textrm{III} radio bursts, 
   while those propagating downward produce HXR emissions and drive chromospheric condensation observed in the Si {\sc iv} line. 
   The electrons trapped in the rising filament generate a microwave burst lasting for $\le$40 s. Bidirectional outflows at the base of jet are manifested by significant line broadenings of the Si {\sc iv} line. 
   The blueshifted Doppler velocities of outflows range from -13 to -101 km s$^{-1}$. The redshifted Doppler velocities of outflows range from $\sim$17 to $\sim$170 km s$^{-1}$.}
   {Our multiwavelength observations of the flare-related jet are in favor of the breakout jet model and are important for understanding the acceleration and transport of nonthermal electrons.}


\maketitle

\section{Introduction} \label{s-intro}
Jet-like activities as a result of magnetic reconnection are ubiquitous in the solar atmosphere. Small-scale jets with lower energy budgets and shorter lifetimes include spicules \citep{dep04,sam19}, 
chromospheric jets \citep{shi07,liu09,tian14a}, bidirectional plasma jets related to explosive events \citep{inn97,li18a}, coronal nanojets \citep{ant20} and mini-jets \citep{chen20b}. 
Large-scale jets with higher energy budgets and longer lifetimes include H$\alpha$ surges \citep{sch95,liu04} and coronal jets \citep{cir07,sav07,sm07,rao16,shen21}.
Coronal jets are transient collimated outflows propagating along open magnetic field or large-scale closed loops \citep{shi92,shi94,zqm12,huang12,huang20}. 
The speeds of jets can reach hundreds of km s$^{-1}$ \citep{sm96,cul07,lu19}. The presence of open field facilitates the escape of electron beams 
accelerated by reconnection at the jet base and the generation of type \textrm{III} radio bursts \citep[e.g.,][]{kru11,gle12,gle18}.

Although the morphology of jets varies from case to case, most of them show an inverse-Y shape or a two-sided shape \citep{mo10,shen19}.
The temperature of a jet decreases from $\sim$10 MK at the base \citep{sm00,chi08,bain09} to a few MK along the spire \citep{zqm14b}.
Sometimes, the hot, fast extreme ultraviolet (EUV) jet is adjacent to or mixed with the cool, slow H$\alpha$ jet \citep{shen17,sak17,sak18,hou20}.
The electron densities of jets range from $\ge$10$^8$ cm$^{-3}$ \citep{you14} to $\ge$10$^{10}$ cm$^{-3}$ \citep{mu17}.
Bright and compact blobs (or plasmoids) are discovered in coronal jets \citep{zqm14b,li19,zn19,jos20a}, 
which are mainly interpreted by magnetic islands as a result of tearing-mode instability in the current sheet \citep{wyp16,ni17}.
Recurring jets produced by successive energy release at the same region are common \citep{chi08,hong19,lu19}. 
Aside from the radial motion, untwisting motions have been detected in helical jets, implying rapid release and transfer of magnetic helicity \citep{chen12,zqm14a,che15,doy19,jos20b}. 

In spite of substantial investigations on jets using multiwavelength observational data, the triggering mechanism is an important issue that needs to be addressed.
Till now, several mechanisms have been proposed, such as magnetic flux emergence and reconnection with pre-existing magnetic fields \citep{yoko96,arc13}, 
magnetic cancellation \citep{pan16,ster17}, minifilament eruption \citep{ster15,hong17,li18b}, and photospheric rotation \citep{par09}.
\citet{wyp17} performed three-dimensional (3D) magnetohydrodynamics (MHD) numerical simulations of a coronal jet driven by filament ejection, 
whereby a region of strongly sheared magnetic field near the solar surface becomes unstable and erupts. 
The authors concluded that energy is initially released via magnetic reconnection at the thin breakout current sheet above the flux rope, 
which is followed by continuing energy release at the thin flare current sheet beneath the erupting filament (or flux rope).
The kinetic evolution of the jet is apparently divided into a slow rise phase before the flux rope opens up and a fast rise phase after the rope totally opens up, 
which correspond to magnetic reconnections at the breakout current sheet and flare current sheet, respectively \citep[][see their Fig. 5]{wyp18}. The breakout jet model is verified observationally 
by the signatures of a rotating jet and fast degradation of the circular flare ribbon following the coherent brightenings of the ribbon associated with the jet \citep{zqm20}.
Breakout reconnection at the null point of a fan-spine structure is recently evidenced by the bidirectional outflows ejected from the reconnection site \citep{yang20}. 
However, reconnection at the flare current sheet below the jet has not been noticed.

In this paper, we report our multiwavelengths of a coronal jet related to a C3.4 circular-ribbon flare that induced simultaneous transverse oscillations of a coronal loop and a filament \citep{zhang20}.
The flare occurred in NOAA active region (AR) 12434 where a series of homologous flares were produced \citep{zqm16a,zqm16b}.
This paper is arranged as follows. Data analysis is described in Sect.~\ref{s-data}. The results are presented in Sect.~\ref{s-res} and compared with previous works in Sect.~\ref{s-disc}.
Finally, a brief summary is given in Sect.~\ref{s-sum}.

\section{Observations and data analysis} \label{s-data}
The C3.4 flare and jet were observed by the Atmospheric Imaging Assembly \citep[AIA;][]{lem12} on board the Solar Dynamics Observatory (SDO) on 2015 October 16.
SDO/AIA takes full-disk images in two ultraviolet (UV; 1600 and 1700 {\AA}) and seven EUV (94, 131, 171, 193, 211, 304, and 335 {\AA}) wavelengths.
The line-of-sight (LOS) magnetograms of the photosphere were observed by the Helioseismic and Magnetic Imager \citep[HMI;][]{sch12} on board SDO.
The level\_1 data of AIA and HMI were calibrated using the standard solar software (SSW) program \texttt{aia\_prep.pro} and \texttt{hmi\_prep.pro}, respectively.

The flare and jet were also observed by the Interface Region Imaging Spectrograph \citep[IRIS;][]{dep14} Slit-Jaw Imager (SJI) in 1330 {\AA} ($\log T\approx4.4$) 
and 1400 {\AA} ($\log T\approx4.8$) with a field of view (FOV) of 134$\arcsec\times$129$\arcsec$. 
Spectroscopic observation of the flare was in the ``large coarse 8-step raster'' mode using four spectral windows (C {\sc ii}, Mg {\sc ii}, O {\sc i}, and Si {\sc iv}).
Each raster had 8 steps from east to west and covered an area of 14$\arcsec\times$129$\arcsec$. The step cadence and exposure time were $\sim$10 s and 8 s.  
The spectra were preprocessed using the standard SSW programs \texttt{iris\_orbitvar\_corr\_l2.pro} and \texttt{iris\_prep\_despike.pro}. 
The Si {\sc iv} 1402.80 {\AA} line ($\log T\approx4.8$) is optically thin and can be fitted by single or multicomponent Gaussian functions.
The reference line center of Si {\sc iv} is set to be 1402.80 {\AA} \citep{li15a,yu20a}. The uncertainty in Doppler velocity is $\sim$2 km s$^{-1}$.

Soft X-ray (SXR) light curves of the flare in 0.5$-$4 and 1$-$8 {\AA} were recorded by the GOES spacecraft. 
Hard X-ray (HXR) fluxes of the flare at different energy bands were obtained from observations of the Reuven Ramaty High-Energy 
Solar Spectroscopic Imager \citep[RHESSI;][]{lin02} and the Gamma-ray Burst Monitor \citep[GBM;][]{mee09} on board the Fermi spacecraft.
The time cadence of Fermi/GBM switched from an ordinary value of 0.256 s before flare to 0.064 s during the flare.
Radio dynamic spectra of the flare were recorded by the ground-based stations belonging to the e-Callisto network\footnote{http://www.e-callisto.org}.
The observational parameters are listed in Table~\ref{tab-1}.

To obtain the 3D magnetic configuration of the AR before flare, we utilize the ``weighted optimization'' method \citep{wie04,wie12} to perform a nonlinear force-free field (NLFFF) 
extrapolation based on the photospheric vector magnetograms observed by HMI at 08:48 UT. The azimuthal component of the inverted vector magnetic field 
was processed to correct the 180$\degr$ ambiguity \citep{leka09}. The vector field in the image plane was transformed to the heliographic plane \citep{gary90}.
The extrapolation was carried out within a box of 380$\times$400$\times$512 uniformly spaced grid points with $dx=dy=dz=1\farcs2$. 
The squashing factor $Q$ \citep{dem96,tit02} and twist number $\mathcal{T}_w$ \citep{ber06} were calculated with the code developed by \citet{liu16}.

\begin{table}
\centering
\caption{Description of the observational parameters.}
\label{tab-1}
\begin{tabular}{lcccc}
\hline\hline
Instru. & $\lambda$   & Time &  Cad. & Pix. Size \\ 
                  & ({\AA})         &  (UT) &  (s)           & (\arcsec) \\
\hline
AIA & 94$-$304 & 08:50-09:18 &  12 & 0.6 \\
AIA & 1600        & 08:50-09:18 &  24 & 0.6 \\
HMI & 6173       & 08:48-09:18 & 45 & 0.6 \\
SJI   & 1330       & 08:50-09:18 & 9, 10 & 0.166 \\
SJI   & 1400       & 08:50-09:18 & 19, 20 & 0.166 \\
GOES     & 0.5$-$4 & 08:58-09:18 &  2.05 & ... \\
GOES     & 1$-$8      &  08:58-09:18 &  2.05 & ... \\
RHESSI  & 6$-$50 keV & 09:00-09:10  &  4.0   & ... \\
GBM       & 4$-$50 keV & 08:58-09:10  &  0.064 & ... \\
KRIM      & 250$-$350 MHz & 08:58-09:01 & 0.25 & ... \\
blen5m   & 0.98$-$1.27 GHz & 09:00-09:02 & 0.25 & ... \\
ZSTS     & 1.41$-$1.43 GHz  & 09:00-09:02 & 0.25 & ... \\
\hline
\end{tabular}
\end{table}

\begin{figure}
\includegraphics[width=9cm,clip=]{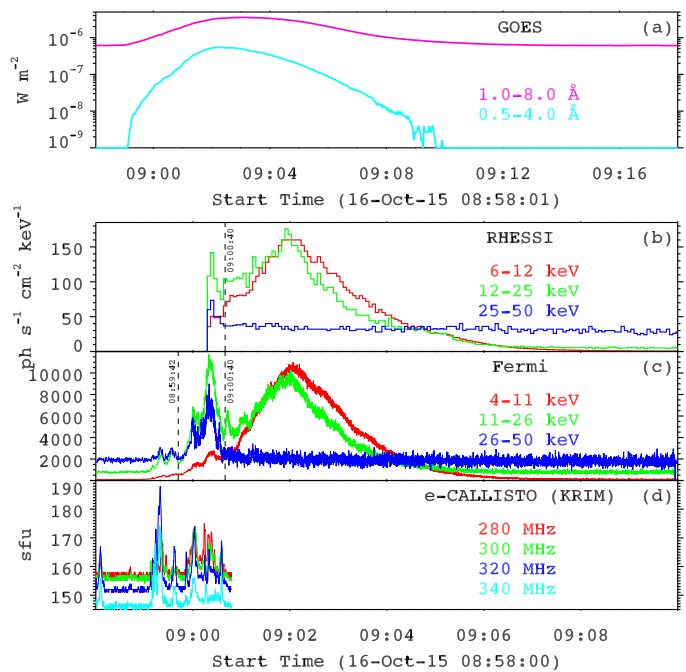}
\centering
\caption{(a) SXR light curves of the C3.4 flare in 0.5$-$4 {\AA} (cyan line) and 1$-$8 {\AA} (magenta line).
(b-c) HXR light curves recorded by RHESSI and Fermi at different energy bands.
(d) Radio light curves recorded by e-Callisto/KRIM at 280, 300, 320, and 340 MHz.}
\label{fig1}
\end{figure}

\begin{figure*}
\includegraphics[width=16cm,clip=]{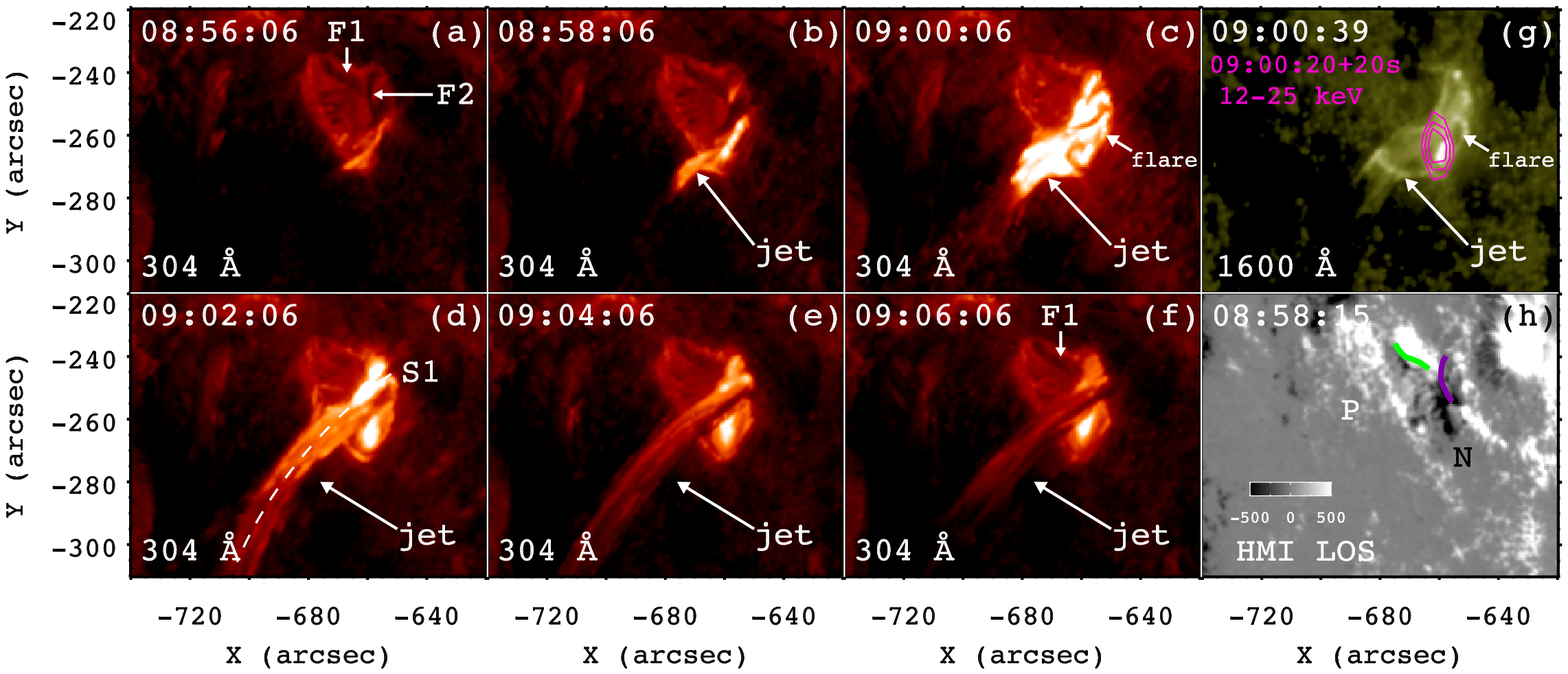}
\centering
\caption{(a-f) Snapshots of the AIA 304 {\AA} images. The arrows point to two minifilaments (F1 and F2), flare, and jet.
In panel (d), the curved slice (S1) is used to investigate the radial propagation of the jet.
(g) AIA 1600 {\AA} image at 09:00:39 UT. The intensity contours of the HXR image are superposed with magenta lines.
(h) HMI LOS magnetogram at 08:58:15 UT. The thick green and purple lines represent F1 and F2 in panel (a).
The whole evolution of the event observed in 304 {\AA} is shown in a movie (\textit{anim304.mov}) available online.}
\label{fig2}
\end{figure*}

\section{Results} \label{s-res}
\subsection{Minifilament eruption and circular-ribbon flare} \label{s-flare}
Figure~\ref{fig1}(a) shows the SXR light curves of the flare. 
The SXR emissions started to rise at $\sim$08:58 UT and reached peak values at $\sim$09:03 UT before declining gradually until $\sim$09:12 UT. 
In Fig.~\ref{fig2}, the EUV images observed by AIA in 304 {\AA} illustrate the whole evolution of the event (see also the online movie \textit{anim304.mov}). 
Before flare, two short minifilaments (F1 and F2) resided in the AR (panel (a)). With the slow rising of F2, 
the EUV intensities of the flare started to increase at $\sim$08:58 UT and reached peak values at $\sim$09:00 UT (panels (b-c)). 
Meanwhile, a coronal jet propagates in the southeast direction, which was also observed in 1600 {\AA} (panel (g)).
The F1 close to F2, however, was undisturbed and survived the flare (panel (f)).
Both of the minifilaments were located along the polarity inversion lines (panel (h)).
The morphology and evolution of the C3.4 flare were quite analogous to those of C4.2 flare starting at $\sim$13:36 UT \citep{zqm16a,dai20}.

HXR light curves of the flare recorded by RHESSI and Fermi at different energy bands are plotted in Fig.~\ref{fig1}(b-c). 
Note that there was no observation from RHESSI until 09:00:20 UT. Before 08:59:42 UT, there were two small peaks (panel (c)). 
The HXR emissions at 11$-$26 keV and 26$-$50 keV started to increase sharply at 08:59:42 UT and peaked at 09:00:20 UT before decreasing to $\sim$09:00:40 UT.
The HXR peaks indicate that the most drastic release of energy and particle acceleration took place during 08:59:42$-$09:00:40 UT ($\sim$60 s).
Afterwards, the HXR emissions between 4 and 26 keV increased gradually and peaked around 09:02:00 UT before decreasing slowly until $\sim$09:04:30 UT, 
implying their thermal nature from hot plasma ($\sim$10 MK) as a result of ongoing chromospheric evaporation.

\begin{figure*}
\includegraphics[width=14cm,clip=]{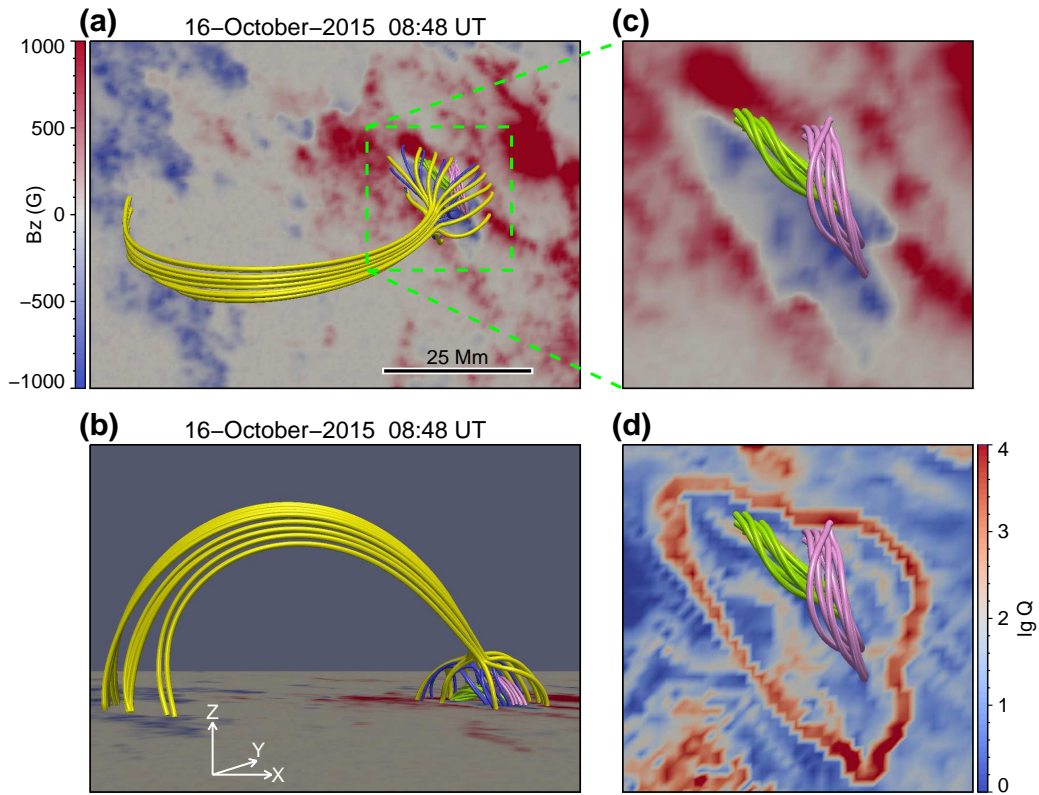}
\centering
\caption{Top view (a) and side view (b) of the 3D magnetic configuration of AR 12434. 
The blue and yellow lines represent the fan-spine field lines.
The green and light violet lines represent the field lines of F1 and F2, respectively.
(c) Close-up of the flare region within the dashed box of panel (a).
(d) Spatial distribution of $\log Q$ at $z=0$ within the flare region, which is overlapped with the field lines of F1 and F2.
The 3D magnetic configuration from different perspectives is shown in a movie (\textit{anim3d.mov}) available online.}
\label{fig3}
\end{figure*}

Figure~\ref{fig2}(h) shows the LOS magnetogram of the flare region observed by HMI at 08:58:15 UT, featuring a central negative polarity surrounded by positive polarities.
Such a magnetic pattern is indicative of the fan-spine topology in the corona \citep[e.g.,][]{zqm12,li18b,hou19,yang20}. 
In Fig.~\ref{fig3}, the left panels demonstrate the top and side views of 3D magnetic configuration of AR 12434.
The fan-spine topology is clearly depicted by the blue and yellow lines, and the outer spine is connected to a remote negative polarity to the southeast of fan dome.
Below the dome, the green and light violet lines represent the field lines of F1 and F2, respectively. A close-up of the flare region is displayed in Fig.~\ref{fig2}(c). 
The magnetic fields supporting the two minifilaments are sheared arcades instead of twisted flux ropes.
The $\mathcal{T}_w$ lies in the range of 0.45$-$0.65 for the lower F1 and 0.6$-$0.7 for the upper F2.
Figure~\ref{fig3}(d) shows the spatial distribution of $\log Q$ within the flare region, 
highlighting the closed ribbon of high $Q$, which is excellently cospatial with the bright circular ribbon (Fig.~\ref{fig2}(c)). 

\begin{figure}
\includegraphics[width=8cm,clip=]{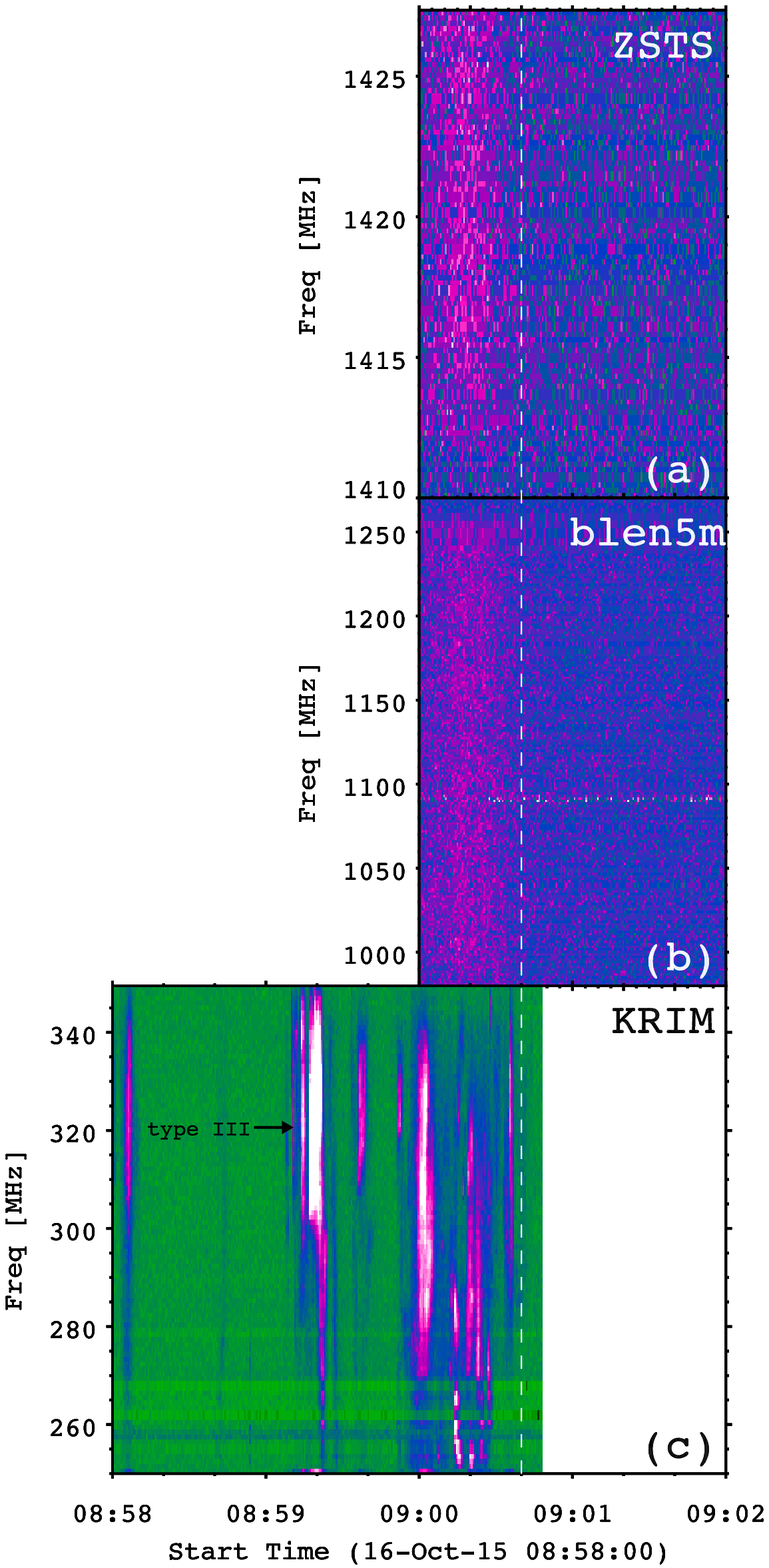}
\centering
\caption{Radio dynamic spectra of the flare recorded by three ground-based stations of e-Callisto network.
The white dashed line denotes the time at 09:00:40 UT. In panel (c), the black arrow points to the type \textrm{III} bursts with fast frequency drift rates.}
\label{fig4}
\end{figure}

The flare was accompanied by type \textrm{III} radio bursts at 250$-$350 MHz. In Fig.~\ref{fig4}, the bottom panel shows the radio dynamic spectra of the flare recorded by e-Callisto/KRIM.
The bursts with fast frequency drift rates are noticeable during 08:59:10$-$09:00:40 UT. The radio fluxes at 280, 300, 320, and 340 MHz are extracted and plotted with colored lines in Fig.~\ref{fig1}(d).
It is revealed that the peaks in radio are roughly correlated with the peaks in HXR, suggesting their common origin.
The reconnection-accelerated electrons propagate downward to produce HXR emissions in the chromosphere 
and propagate upward along open field to produce type \textrm{III} bursts at the same time \citep{zqm16b}.
Combining with the results of extrapolation in Fig.~\ref{fig3}, the real magnetic configuration of the flare region is consistent with previous schematic illustration \citep[][see their Fig. 1]{wang12}.

\subsection{Coronal jet} \label{s-jet}
\begin{figure}
\includegraphics[width=8cm,clip=]{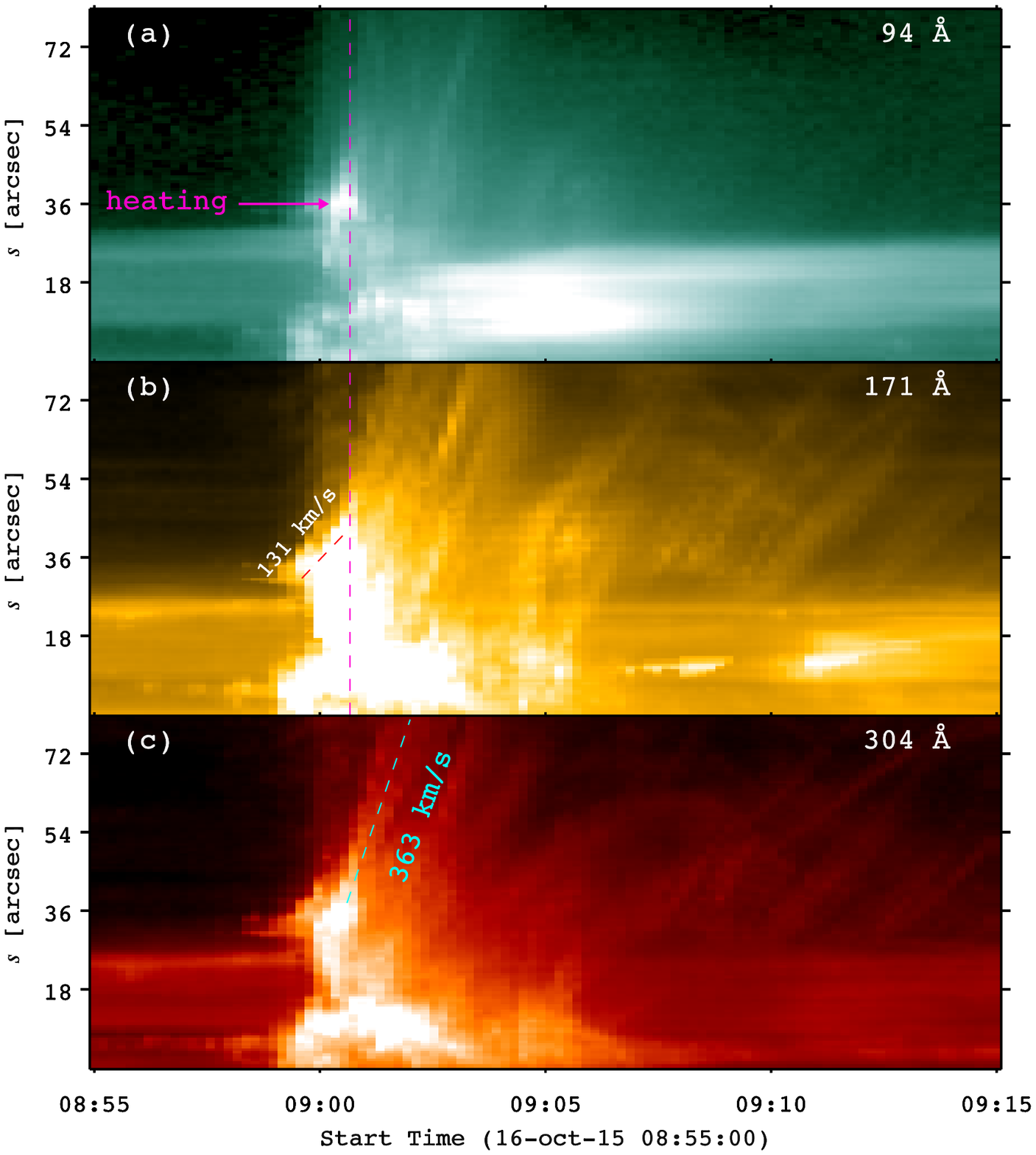}
\centering
\caption{Time-distance diagrams of S1 in 94, 171, and 304 {\AA}. The magenta dashed line denotes the time at 09:00:40 UT.
The plane-of-sky speeds of the jet during the slow rise ($\sim$131 km s$^{-1}$) and fast rise ($\sim$363 km s$^{-1}$) are labeled.
On the $y$-axis, $s=0$ and $s=80\farcs4$ denote the northwest and southeast endpoints of S1, respectively.}
\label{fig5}
\end{figure}

To investigate the radial propagation of the jet, an artificial slice (S1) along the jet spire is selected in Fig.~\ref{fig2}(d), which is $\sim$80$\arcsec$ in length. 
Time-distance diagrams of S1 in 94 {\AA} ($\log T\approx6.8$), 171 {\AA} ($\log T\approx5.8$), and 304 {\AA} ($\log T\approx4.7$) are displayed in Fig.~\ref{fig5}.
The jet is distinctly observed in all EUV wavelengths, indicating its multithermal nature \citep{zqm14b,jos20a}.
The kinetic evolution is divided into two phases: a slow rise phase at a plane-of-sky speed of $\sim$131 km s$^{-1}$ and a fast rise phase at a speed of $\sim$363 km s$^{-1}$, respectively.
The transition between the two phases occurred at $\sim$09:00:40 UT, which is denoted by the magenta dashed line.
Impulsive heating to reach a temperature of $\ge$6 MK is concurrent with the turning point (panel (a)).
The evolution of jet is basically consistent with the breakout model \citep{wyp17,wyp18}. 
Magnetic reconnections at the breakout current sheet and flare current sheet lead to intense electron acceleration and heating.

\begin{figure}
\includegraphics[width=9cm,clip=]{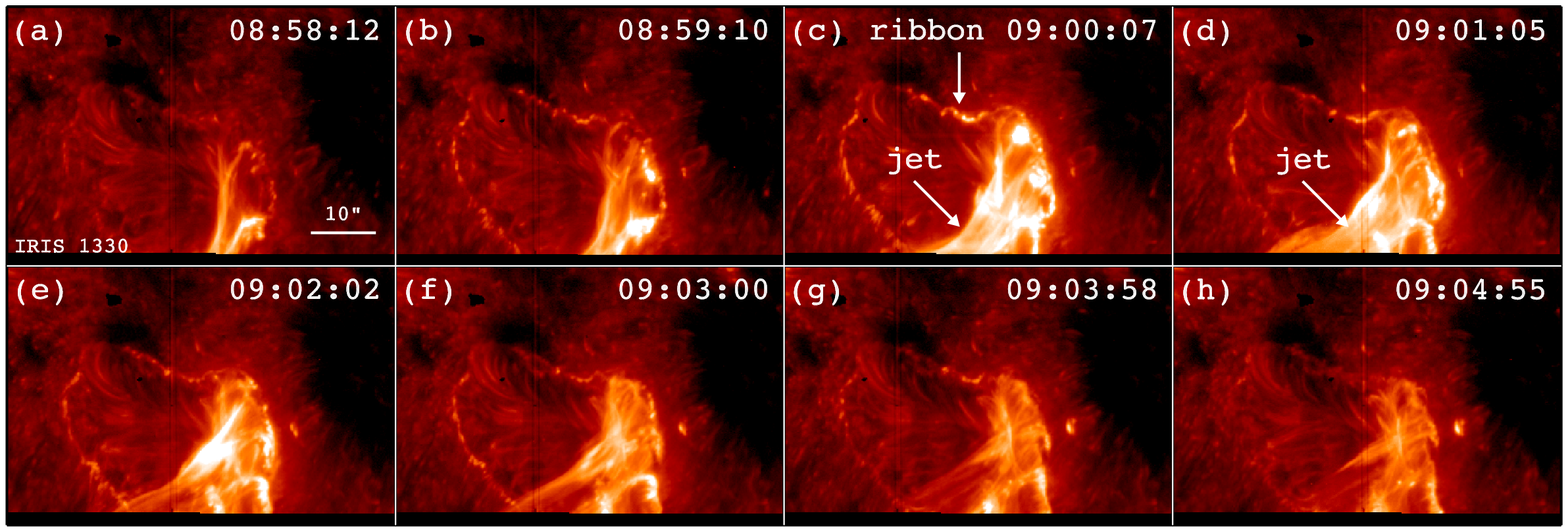}
\centering
\caption{Snapshots of the IRIS/SJI 1330 {\AA} images with a FOV of 60$\arcsec\times$40$\arcsec$. The arrows point to the bright circular ribbon and jet.
The evolution of the event observed in 1330 {\AA} is shown in a movie (\textit{anim1330.mov}) available online.}
\label{fig6}
\end{figure}

\begin{figure}
\includegraphics[width=8cm,clip=]{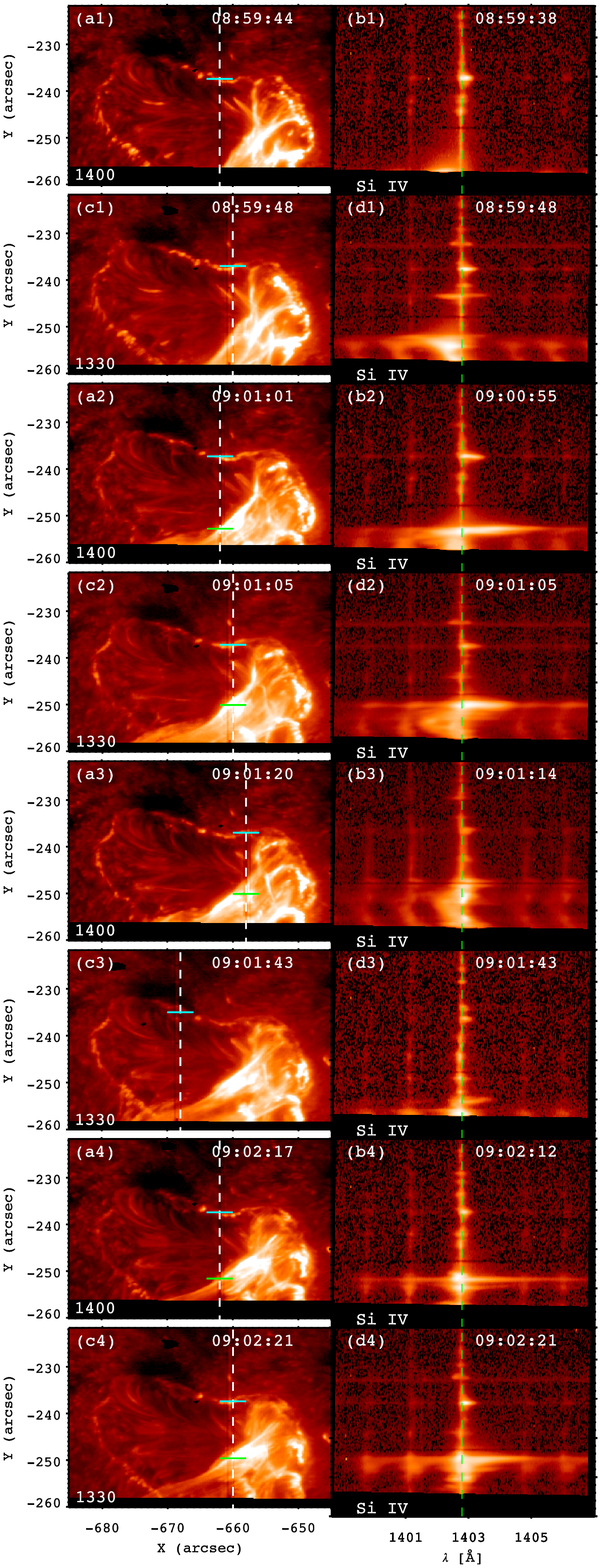}
\centering
\caption{Selected FUV images observed by SJI (left panels) and the corresponding Si {\sc iv} spectra of the slit (right panels).
The white dashed lines represent the slit positions of raster observations. The short cyan lines mark the positions of jet for spectral fittings in Fig.~\ref{fig8}.
The green dashed lines represent the reference line center of Si {\sc iv}, i.e., 1402.80 {\AA}.}
\label{fig7}
\end{figure}

In Fig.~\ref{fig6}, the 1330 {\AA} images observed by IRIS/SJI illustrate the evolution of flare and jet in FUV wavelength, which is similar to that in EUV wavelengths.
Coherent brightenings of the circular ribbon took place around 09:00:07 UT, indicating null point reconnection (see panel (c) and the online movie \textit{anim1330.mov}).
Unfortunately, the southern part of flare and major part of jet spire were not observed due to the limited FOV of IRIS.
The raster observations enable us to carry out spectral analysis of the flare and jet base.
In Fig.~\ref{fig7}, the left column shows selected FUV images observed by SJI when the jet was covered by the slit.
The right column shows the corresponding Si {\sc iv} spectra of the slit.
The spectra of the jet are generally blueshifted, meaning that the jet materials are moving toward the observer.

\begin{figure}
\includegraphics[width=9cm,clip=]{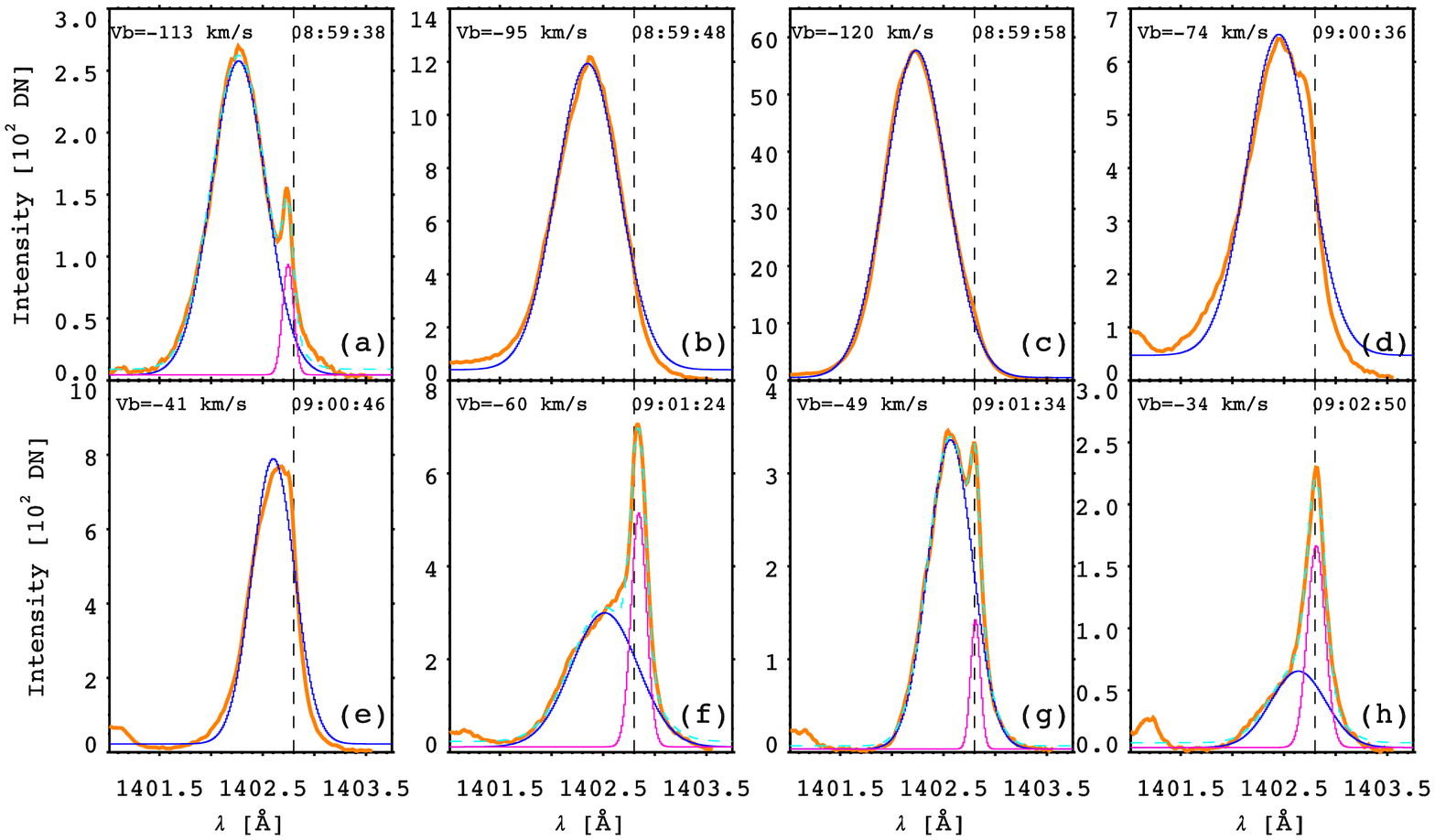}
\centering
\caption{Line profiles of Si {\sc iv} at the positions of slit, marked with short cyan lines in Fig.~\ref{fig7}. 
The orange lines represent the observed profiles. The magenta and blue lines stand for the fitted components.
The blueshifted Doppler velocities ($v_b$) of the jet are labeled.}
\label{fig8}
\end{figure}

To precisely quantify the Doppler velocities of the jet, the line profiles of Si {\sc iv} at the positions marked with cyan lines are extracted and plotted with orange lines in Fig.~\ref{fig8}.
The entirely blueshifted profiles are satisfactorily fitted with a single-Gaussian function (blue lines in panels (b-e)), and the calculated Doppler velocities ($v_b$) are labeled on top of each panel.
The profiles with an enhanced blue wing are fitted with a double-Gaussian function (blue and magenta lines in the remaining panels). 
The sum of two components are drawn with cyan dashed lines, which nicely agree with the observed profiles (orange lines), meaning that the results of double-Gaussian fitting are acceptable.
The calculated Doppler velocities ($v_b$) of the blueshifted component are also labeled on top of each panel. 
In Fig.~\ref{fig12}, the values of $v_b$, ranging from -34 to -120 km s$^{-1}$, are marked with blue circles.

\subsection{Chromospheric condensation} \label{s-cc}
\begin{figure}
\includegraphics[width=8cm,clip=]{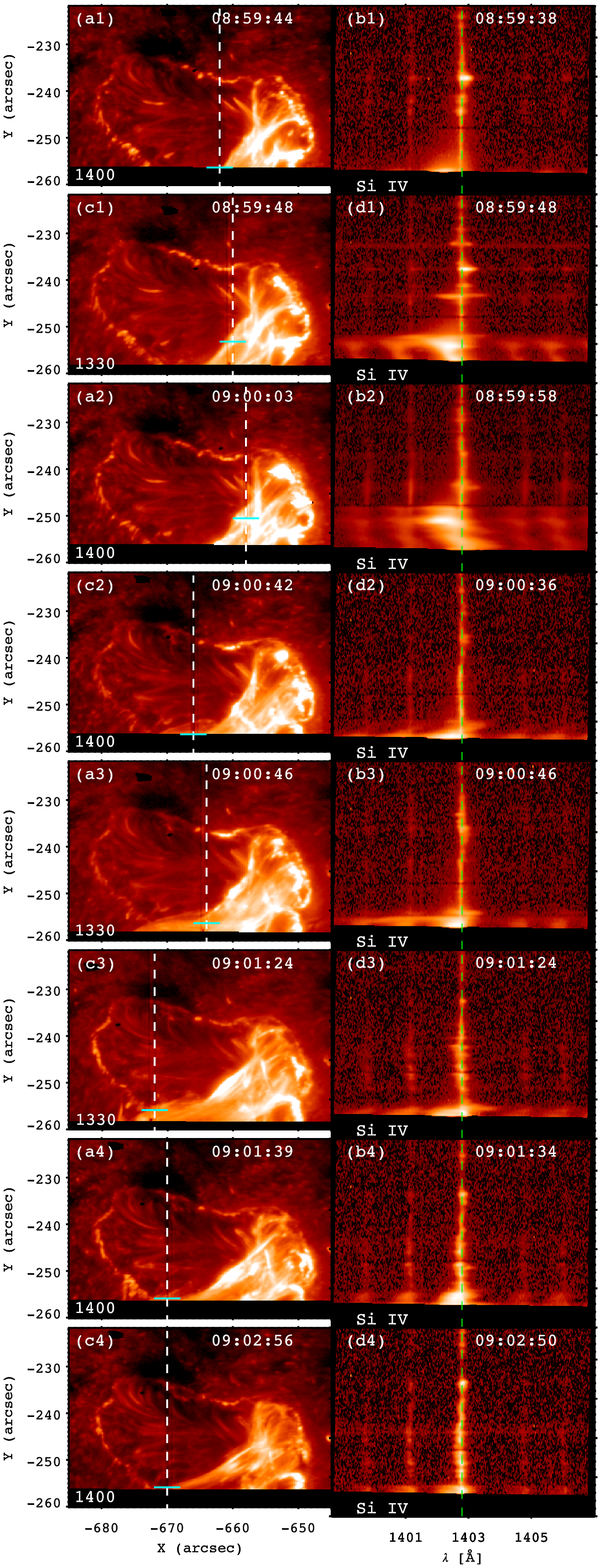}
\centering
\caption{Selected FUV images observed by SJI (left panels) and the corresponding Si {\sc iv} spectra of the slit (right panels).
The white dashed lines represent the slit positions of raster observations. The short cyan lines mark the positions of ribbon for spectral fittings in Fig.~\ref{fig10}.
In panels (a2), (a3), (a4), (c2), and (c4), the short green lines  mark the positions of jet base for spectral fittings in Fig.~\ref{fig11}.
The green dashed lines represent the reference line center of Si {\sc iv}.}
\label{fig9}
\end{figure}

Chromospheric condensation at circular flare ribbons has been observed and investigated in the homologous C3.1 and C4.2 flares \citep{zqm16a,zqm16b}. 
The redshifted velocities of the downflow reach up to $\sim$60 km s$^{-1}$ using the observations of Si {\sc iv} line.
The condensation is primarily driven by reconnection-accelerated nonthermal electrons \citep{li15b,li17}.
In Fig.~\ref{fig9}, likewise, the left column shows selected FUV images observed by SJI when the circular ribbon was covered by the slit.
The right column shows the corresponding Si {\sc iv} spectra of the slit.
The spectra of ribbon are redshifted, indicating plasma downflow or chromospheric condensation.

\begin{figure}
\includegraphics[width=9cm,clip=]{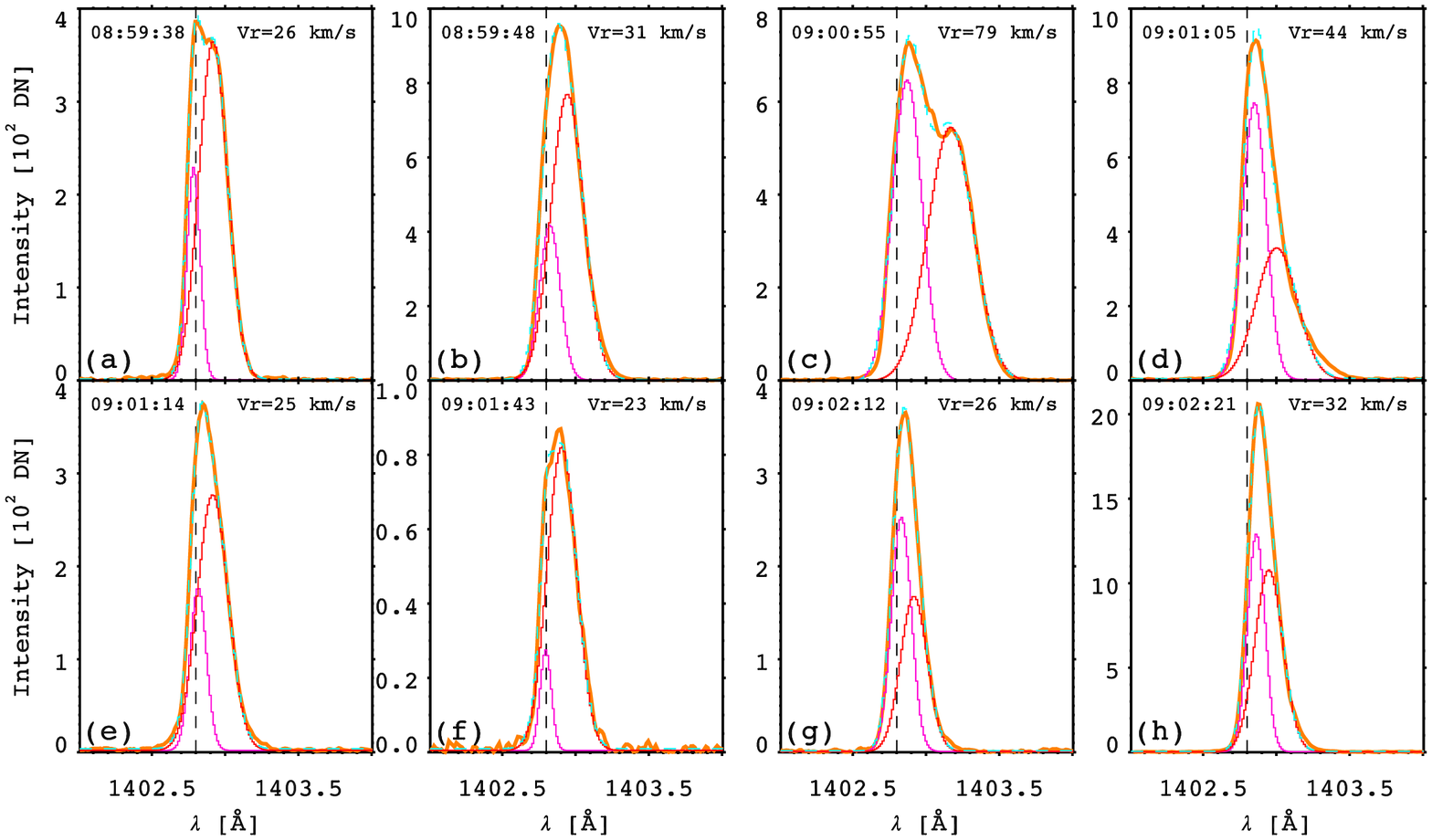}
\centering
\caption{Line profiles of Si {\sc iv} at the positions of slit, marked with short cyan lines in Fig.~\ref{fig9}. 
The orange lines represent the observed profiles. The magenta and red lines stand for the fitted components.
The redshifted Doppler velocities ($v_r$) of chromospheric condensation are labeled.}
\label{fig10}
\end{figure}

To quantify the Doppler velocities of condensation, the line profiles of Si {\sc iv} at the positions marked with cyan lines are extracted and plotted with orange lines in Fig.~\ref{fig10}.
The profiles are fitted with a double-Gaussian function (red and magenta lines), and the sum of two components are drawn with cyan dashed lines, 
which gratifyingly agree with the observed profiles. The calculated Doppler velocities ($v_r$) of the redshifted component are labeled on top of each panel.
In Fig.~\ref{fig12}, the values of $v_r$, ranging from $\sim$20 to $\sim$80 km s$^{-1}$, are marked with red circles. The cause of condensation will be discussed in Sect.~\ref{s-cause}.

\subsection{Bidirectional outflows} \label{s-bf}
\begin{figure}
\includegraphics[width=9cm,clip=]{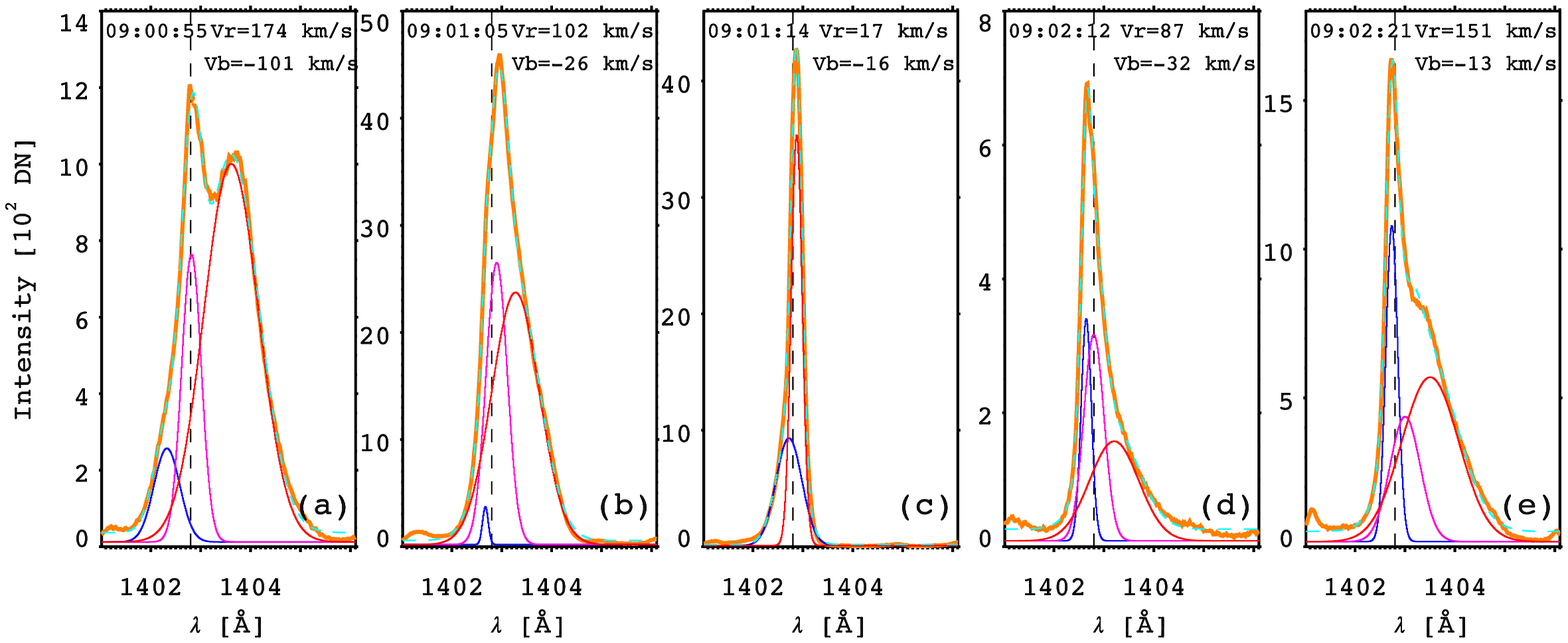}
\centering
\caption{Line profiles of Si {\sc iv} at the positions of slit, marked with short green lines in Fig.~\ref{fig9}.
The orange lines represent the observed profiles. The blue, magenta, and red lines stand for the fitted components.
The redshifted ($v_r$) and blueshifted ($v_b$) Doppler velocities of bidirectional reconnection outflows are labeled.}
\label{fig11}
\end{figure}

\begin{figure}
\includegraphics[width=8cm,clip=]{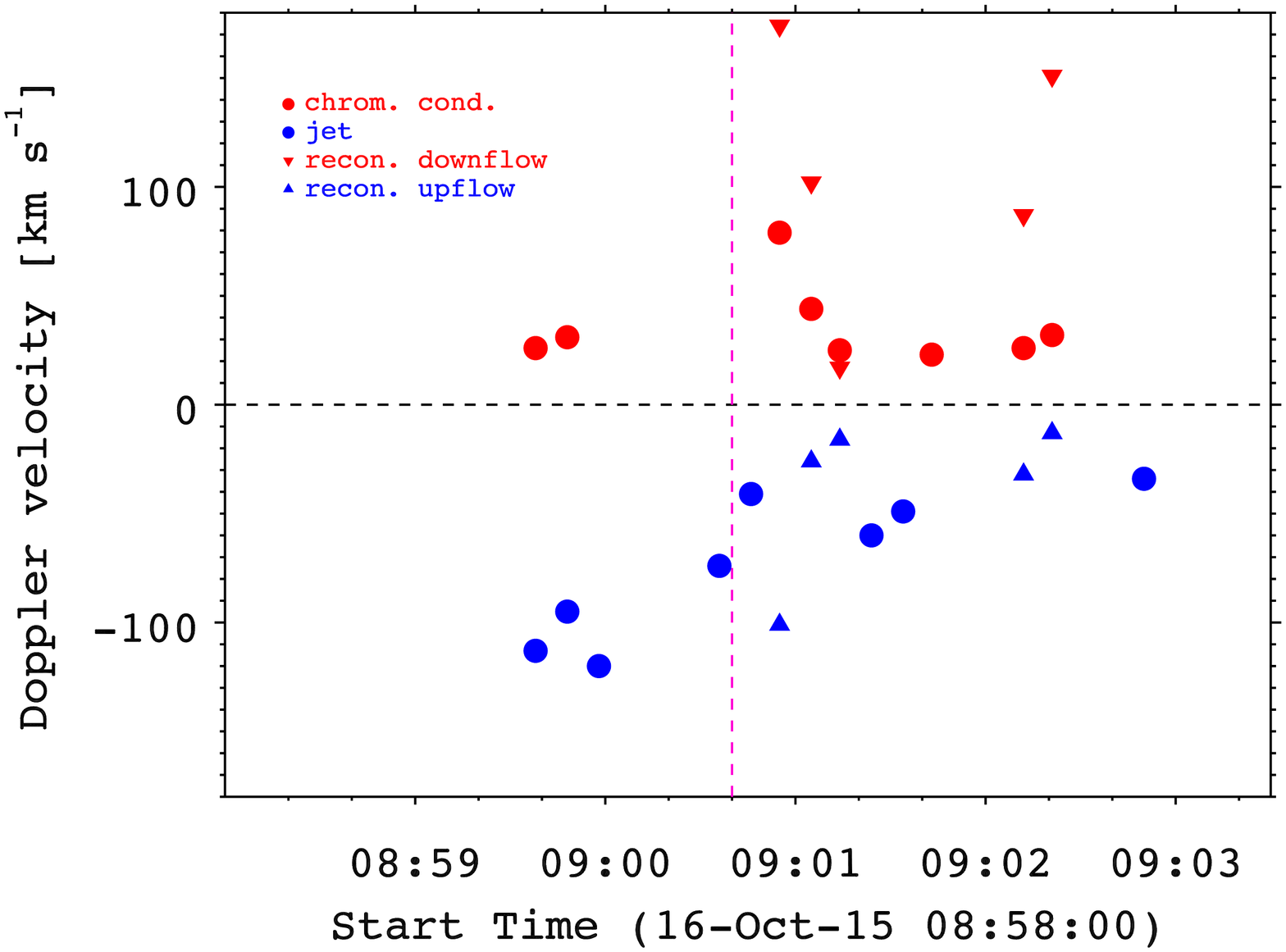}
\centering
\caption{Time evolutions of the Doppler velocities for jet (blue circles), chromospheric condensation (red circles), 
reconnection downflow (red triangles) and upflow (blue triangles), respectively.}
\label{fig12}
\end{figure}

As mentioned before, magnetic reconnection occurs not only at the breakout current sheet above the eruptive filament (or flux rope) but also at the flare current sheet below the flux rope \citep{wyp17,wyp18}.
In Fig.~\ref{fig9}, dramatic line broadenings of the Si {\sc iv} line at the jet base are shown in panels (b2), (b3), (b4), (d2), and (d4), 
implying simultaneous bidirectional reconnection outflows below the jet \citep[e.g.,][]{tian14b,li18c,xue18}.
To work out the Doppler velocities of outflows, the line profiles of Si {\sc iv} at the positions marked with short green lines are extracted and plotted with orange lines in Fig.~\ref{fig11}.
The profiles are fitted with a triple-Gaussian function (blue, magenta, and red lines) except in panel (c), which is fitted with a double-Gaussian function.
The sum of multiple components are drawn with cyan dashed lines.
The calculated Doppler velocities of the redshifted component ($v_r$) and blueshifted component ($v_b$) are labeled on top of each panel.
In Fig.~\ref{fig12}, the velocities of reconnection upflow, in the range of -13 and -101 km s$^{-1}$, are marked with blue triangles.
The velocities of reconnection downflow, in the range of $\sim$17 to $\sim$170 km s$^{-1}$, are marked with red triangles.
We note that the fast reconnection outflows are observed after 09:00:40 UT, when the jet is rising quickly at a speed of $\sim$363 km s$^{-1}$.

It should be emphasized that the chromospheric condensation takes place at the circular ribbon as marked by short cyan lines in Fig.~\ref{fig9}.
Only redshifted downflow at speeds of 20$-$80 km s$^{-1}$ could be identified in the spectra (Figs.~\ref{fig9}, \ref{fig10}).
The bidirectional outflows are observed inside the circular ribbon and close to the jet base.
Simultaneous upflow and downflow could be recognized in the spectra (Figs.~\ref{fig9}, \ref{fig11}) 
and the Doppler velocities of bidirectional outflows are significantly higher than those of condensation.
These are the main differences between chromospheric condensation and bidirectional outflows.

\section{Discussion} \label{s-disc}
\subsection{Magnetic reconnection} \label{s-mr}
Magnetic reconnection is believed to play a key role in the energy release of solar flares \citep{pri02,pri09}.
Direct evidences of reconnection at current sheets are abundant, including the bidirectional inflows and outflows \citep[e.g.,][]{sav12,ning14,wu16,xue18,yan18,chen20a,yu20b}, 
change of magnetic topology \citep{su13}, localized heating to $\ge$10 MK \citep{sea17,li18c,war18}, and creation of magnetic islands \citep{li16}.
Using 3D MHD numerical simulations, \citet{wyp17} proposed a universal model for solar eruptions, including eruptive flares and coronal jets.
Breakout reconnection around an X-type null point in a quadrupolar magnetic configuration is observed and investigated by \citet{chen16}.
Observations of fast reconnection at the breakout current sheet of a fan-spine structure are carried out by \citet{yang20}.

In our study, impulsive interchange reconnection at the null point was manifested by coherent brightenings of the circular ribbon, HXR peaks, and type \textrm{III} radio bursts 
around 09:00:00 UT (Fig.~\ref{fig1}(c-d) and Fig.~\ref{fig6}(c)). Line broadening as a result of bidirectional outflows was absent before the minifilament broke through the fan surface.
During the reconnection at the flare current sheet underneath the filament, pronounced upflows and downflows at the jet base were evidenced by significant Doppler line broadenings of Si {\sc iv}.
The jet did not show untwisting motion during its radial propagation, which is probably due to the small $\mathcal{T}_w$ of the sheared arcade supporting the minifilament.
Taken as a whole, our multiwavelength observations of the flare-related jet support the breakout jet model \citep{wyp18}.

Using combined observations of a small prominence eruption on 2014 May 1, \citet{ree15} found evidence for reconnection between the prominence magnetic field and the overlying field.
Reconnection outflows at a plane-of-sky speed of $\sim$300 km s$^{-1}$ and Doppler velocity of $\sim$200 km s$^{-1}$ were detected by SDO/AIA and IRIS, respectively.
Moreover, possible reconnection site below the prominence is found (see their Fig. 10). The authors, however, concluded that the reconnection was triggered not by breakout reconnection, 
but by reconnection occurring along and beneath the prominence.

Using multiwavelength observations from SDO, Hinode/XRT, IRIS, and the DST of Hida Observatory, \citet{sak18} analyzed a jet-related C5.4 flare on 2014 November 11. 
The morphology and magnetic configuration of the jet were somewhat similar to the jet in our study. 
However, the C5.4 flare and jet were caused by magnetic reconnection between the emerging magnetic flux of the satellite spots and the pre-existing ambient fields \citep{sak17}.
Part of the cool H$\alpha$ jet experienced secondary acceleration between the trajactories of the H$\alpha$ jet and the hot SXR jet after it had been ejected from the lower atmosphere,
which is explained by magnetic reconnection between the preceding H$\alpha$ jet and the plasmoid in the subsequent SXR jet.
In our case, the circular-ribbon flare was caused by eruption of a minifilament underlying the null point (Figs.~\ref{fig2}, \ref{fig3}).
The fast rise of the jet after $\sim$09:00:40 UT may result from quick ejection of F2 after null-point reconnection \citep{wyp18}.
The reconnection upflow at the flare current sheet may also contribute to the acceleration of jet, as in the case of coronal mass ejections \citep[CMEs;][]{zj04}.

\subsection{Cause of chromospheric condensation} \label{s-cause}
As mentioned above, chromospheric condensation could be driven by electron beam heating \citep[e.g.,][]{li15a,li15b,yu20a}.
During the C4.2 flare on 2015 October 16, explosive chromospheric evaporation took place, which was characterized by plasma upflow at speeds of 35$-$120 km s$^{-1}$ 
in the Fe {\sc xxi} 1354.09 {\AA} line ($\log T\approx7.05$) and downflow at speeds of 10$-$60 km s$^{-1}$ in the Si {\sc iv} 1393.77 {\AA} line \citep{zqm16a}. 
The estimated nonthermal energy flux above 20 keV exceeds the threshold ($\sim$10$^{10}$ erg s$^{-1}$ cm$^{-2}$) for explosive evaporation \citep{fis85}.
Condensation during the C3.1 flare associated with a type \textrm{III} burst is also believed to be driven by electron beams \citep{zqm16b}.

\begin{figure}
\includegraphics[width=8cm,clip=]{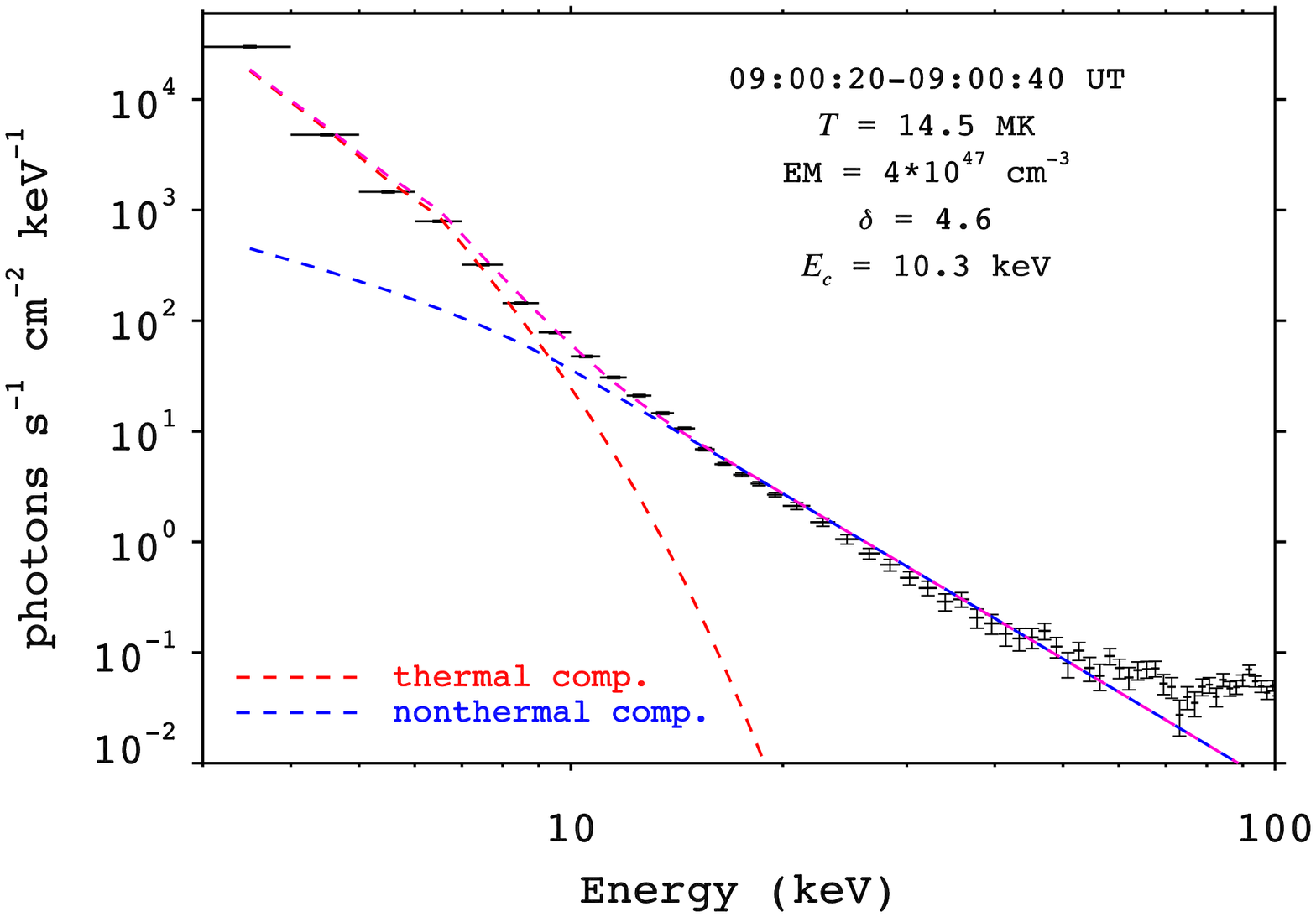}
\centering
\caption{Results of RHESSI spectral fitting during 09:00:20$-$09:00:40 UT. The data points with horizontal and vertical error bars represent the observed data. 
The spectra for thermal and nonthermal component are plotted with red and blue dashed lines, respectively. The sum of two components is shown with a magenta dashed line.
The fitted parameters, including the thermal temperature ($T$), emission measure (EM), electron spectral index ($\delta$), and low-energy cutoff ($E_c$), are labeled.}
\label{fig13}
\end{figure}

To explore the cause of condensation during the C3.4 flare, we focus on nonthermal electrons as usual.
In Fig.~\ref{fig2}(g), the AIA 1600 {\AA} image at 09:00:39 UT is superposed with intensity contours of HXR emission at 12$-$25 keV during 09:00:20$-$09:00:40 UT (magenta lines).
The centroid of the single HXR source is cospatial with the bright inner ribbon, where majority of nonthermal electrons are precipitated.
In Fig.~\ref{fig13}, the HXR spectrum obtained from RHESSI observation is fitted with a thermal component plus a thick-target, nonthermal component consisting of a broken power law \citep{rub15}:
\begin{equation} \label{eqn-1}
F(E)=(\delta-1)\frac{\mathcal{F}_c}{E_c}(\frac{E}{E_c})^{-\delta},
\end{equation}
where $\delta$ and $E_c$ represent the spectral index and cutoff energy of nonthermal electrons. $\mathcal{F}_c=\int_{E_c}^{\infty}F(E)dE$ denotes the total electron flux above $E_c$.
The fitted thermal component is plotted with a red dashed line, with the values of thermal temperature ($T$), emission measure (EM) being labeled.
The fitted nonthermal component is plotted with a blue dashed line, with the values of $\delta$ and $E_c$ being labeled as well.
The sum of two components (magenta dashed line) agrees with the observed spectrum.

Before $\sim$09:00:40 UT, the HXR emissions were predominantly produced by nonthermal electrons (Fig.~\ref{fig1}(c)).
The power of injected electrons above $E_c$ is expressed as:
\begin{equation} \label{eqn-2}
P_{nth}=\int_{E_c}^{\infty}F(E)EdE=\frac{\delta-1}{\delta-2}\mathcal{F}_cE_c.
\end{equation}
Substituting the parameters obtained from HXR spectral fitting in Fig.~\ref{fig13}: $\delta=4.6$, $\mathcal{F}_c=5.6\times10^{35}$ electrons s$^{-1}$, $E_c=10.3$ keV, 
then $P_{nth}$ is estimated to be $\sim$1.3$\times$10$^{28}$ erg s$^{-1}$. 
A lower limit of the total nonthermal energy in electrons is $\sim$2.6$\times$10$^{29}$ erg, since there was no RHESSI observation before 09:00:20 UT. 
Considering that the area of HXR source is in the range of (0.25$-$1)$\times$10$^{18}$ cm$^2$, 
the total nonthermal energy flux is estimated to be (1.3$-$5.2)$\times$10$^{10}$ erg s$^{-1}$ cm$^{-2}$, which is adequate to drive chromospheric condensation at flare ribbon.

After $\sim$09:00:40 UT, the HXR emission at 25$-$50 keV decreased to the pre-eruption level, while the emission at 12$-$25 keV increased gradually with episodic spikes till $\sim$09:02:00 UT.
To investigate the role of heat conduction in driving condensation, we estimate the energy flux of heat conduction using the expression:
\begin{equation} \label{equ-3}
F_{cond}=\kappa_0\frac{T^{7/2}}{L},
\end{equation}
where $\kappa_0\approx10^{-6}$ erg s$^{-1}$ cm$^{-1}$ K$^{-7/2}$, $T\approx14.5$ MK denotes the thermal temperature in the corona, and $L\approx\frac{\pi d}{4}$ denotes the length scale of 
the temperature gradient ($d\approx36\arcsec$ represents the equivalent diameter of circular ribbon). $F_{cond}$ is estimated to be $\sim$5.5$\times$10$^9$ erg s$^{-1}$ cm$^{-2}$.
Therefore, the condensation after $\sim$09:00:40 UT should be driven by a combination of nonthermal electrons and heat conduction \citep{sad15}.

\subsection{Emission mechanism of microwave burst}
As mentioned above, the flare was associated with type \textrm{III} radio bursts, which are generated by electron beams propagating upward along open field \citep{inn11,chen18}.
The top and middle panels of Fig.~\ref{fig4} show radio dynamic spectra recorded by ZSTS and blen5m during 09:00$-$09:02 UT with the same cadence as KRIM.
Enhancement of the microwave emission at 0.98$-$1.43 GHz was obviously demonstrated before 09:00:40 UT.
Contrary to the discrete type \textrm{III} bursts owing to the coherent plasma radiation mechanism \citep{dulk85}, 
the microwave burst seems to be continuous and has no one-to-one correspondence with the type \textrm{III} bursts, 
implying that the microwave burst was not produced by plasma radiation mechanism.
To interpret the relationship between the HXR and radio time profiles of single-loop flares, \citet{kun01} proposed a simple trap model by introducing a critical pitch angle (i.e. loss cone angle).
Those injected high-energy electrons with smaller pitch angles are not reflected by the increasing magnetic field as they approach the loop footpoints and will precipitate on their first approach.
The remaining electrons with pitch angles outside the loss cone will be trapped in the loop unless pitch angle scattering takes place.

In our case, the accelerated nonthermal electrons before $\sim$09:00:40 UT are composed of three groups. The first group propagates along open field to generate 
type \textrm{III} radio bursts (Fig.~\ref{fig4}(c)). The second group precipitates straightforward into the chromosphere and generates HXR emissions (Fig.~\ref{fig1}(b-c)).
The third group is trapped in the rising minifilament and generates microwave burst through the gyrosynchrotron emission mechanism \citep{dulk85,lee00,wu16}.
Of course, the trapped electrons will eventually precipitate once pitch angle scattering switches on.
Existence of trapped nonthermal electrons in a kink-unstable filament undergoing a failed eruption has been reported by \citet{guo12}.
After the minifilament breaks through the null point and totally opens up, there are no trapped electrons any more and the microwave emission vanishes.
This scenario is qualitative and needs to be validated by in-depth investigations. 
Additional cases of microwave bursts like in Fig.~\ref{fig4} have been collected, which will be the topic of our next paper.

\section{Summary} \label{s-sum}
In this work, a coronal jet related to a C3.4 flare in AR 12434 on 2015 October 16 is studied in detail. The main results are summarized as follows:

\begin{enumerate}
\item{Two minifilaments were located under a 3D fan-spine structure before flare. The flare was generated by the eruption of one filament.
The kinetic evolution of the jet was divided into two phases: a slow rise phase at a speed of $\sim$130 km s$^{-1}$ and a fast rise phase at a speed of $\sim$360 km s$^{-1}$ in the plane-of-sky.
The slow rise phase may correspond to breakout reconnection at the breakout current sheet, and the fast rise phase may correspond to reconnection at the flare current sheet.
The transition between the two phases took place at $\sim$09:00:40 UT. The blueshifted Doppler velocities of the jet in the Si {\sc iv} 1402.80 {\AA} line range from -34 to -120 km s$^{-1}$.}
\item{The accelerated high-energy electrons are composed of three groups. Those propagating upward along open field generate type \textrm{III} radio bursts, 
while those propagating downward produce HXR emissions and drive chromospheric condensation. 
The electrons trapped in the rising filament generate a microwave burst lasting for $\le$40 s.}
\item{Bidirectional outflows at the jet base are manifested by significant line broadenings of the Si {\sc iv} line. 
The blueshifted Doppler velocities range from -13 to -101 km s$^{-1}$. The redshifted Doppler velocities range from $\sim$17 to $\sim$170 km s$^{-1}$.
Our multiwavelength observations of the flare-related jet are in favor of the breakout jet model and shed light on the acceleration and transport of nonthermal electrons.}
\end{enumerate}

\begin{acknowledgements}
The authors thank the referee for constructive suggestions and comments.
The authors appreciate Drs. Xiaoli Yan in Yunnan Observatories, Ying Li and Lei Lu in Purple Mountain Observatory, 
Yang Guo in Nanjing University, and Sijie Yu in New Jersey Institute of Technology for valuable discussions.
SDO is a mission of NASA\rq{}s Living With a Star Program. AIA and HMI data are courtesy of the NASA/SDO science teams.
This work is funded by NSFC grants (No. 11773079, 11790302, U1831112, 11903050, 11790304, 11973092, 11573072, and 11703017), 
the International Cooperation and Interchange Program (11961131002), 
the Youth Innovation Promotion Association CAS, CAS Key Laboratory of Solar Activity, National Astronomical Observatories (KLSA202003, KLSA202006), 
and the project supported by the Specialized Research Fund for State Key Laboratories.
\end{acknowledgements}

\end{document}